# Revealing the Hidden Patterns of News Photos: Analysis of Millions of News Photos through GDELT and Deep Learning-based Vision APIs

**Haewoon Kwak    Jisun An**

Qatar Computing Research Institute
Hamad Bin Khalifa University
{hkwak, jan}@qf.org.qa

## Abstract

In this work, we analyze more than two million news photos published in January 2016. We demonstrate i) which objects appear the most in news photos; ii) what the sentiments of news photos are; iii) whether the sentiment of news photos is aligned with the tone of the text; iv) how gender is treated; and v) how differently political candidates are portrayed. To our best knowledge, this is the first large-scale study of news photo contents using deep learning-based vision APIs.

## Introduction

News photos play a crucial role in people's perception of the news (Goldberg 1991). They do not only enhance readers' memory but also deliver emotions otherwise hard to be conveyed (Lester 2013). Regardless of this importance, the innate subtle and nuanced presentation makes it hard to understand news photos quantitatively, compared to well-established text mining techniques that have been applied to analyze textual contents of the news. For example, in order to study how differently men and women are portrayed in news photos, manual coding of photos has been essential, but it is hard to scale.

With advances in deep learning, researchers have a new opportunity to study news photos at scale. The image recognition by deep learning is known to perform even better than humans in many tasks. In 2016, Google announced the beta release of their API for image content analysis, called Google Cloud Vision API, which automatically recognizes objects, faces, and sentiments of a given photo.

In this work, we analyze more than two million news photos published in January 2016. These images are indexed as GDELT Visual Global Knowledge Graph (VGKG) records and undergo image content analysis by Google Cloud Vision API. We demonstrate how well a wide range of research on news photos can be conducted in large-scale by answering the following questions: i) Which objects appear the most in news photos? ii) What are the sentiments of news photos? iii) Is the sentiment of news photos aligned with the tone of the text? iv) How is gender treated? And v) How differently are political candidates portrayed?

We discovered that people appear in 60.3% of CNNs news photos and 53.2% of Fox News news photos while there are some differences in news photos across the news sections (e.g. people delivering a speech in politics and opinion sections). To extend our understanding of photos of people, we analyzed the emotional attributes of news photos. We found that a quarter of all the faces in news photos expressed joy. The number of faces associated with other emotions is negligible. The remaining majority is neutral, without emotional attributes. In addition, by using Microsoft Project Oxford Face API, we found that women smile more and look younger than men in news photos.

We then conducted a case study of assessing the portrayal of Democratic and Republican party presidential candidates in news photos. We found that all the candidates but Sanders had a similar proportion of being labeled as an athlete, which is typically associates with a victory pose or a sharp focus on a face with blurred background. Pro-Clinton media recognized by their endorsements show the same tendency; their Sanders photos are not labeled as an athlete at all. Furthermore, we found that Clinton expresses joy more than Sanders does in the six popular news media. Similarly. pro-Clinton media shows a higher proportion of Clinton expressing joy than Sanders.

## GDELT project

GDELT (Global Data on Events, Location, and Tone) is a recently released open dataset of digital prints of a wide range of media (e.g., news media and blogs). It monitors news media in over 100 languages from the whole world and indexes them (Leetaru and Schrodt 2013).

As they say, "GDELT is an open dataset that attempts to make human society itself computable", but there have been only few studies meeting this expectation. Kwak and An (2014) have conducted a large-scale study on foreign news coverage. While most previous work had to focus on a single or few countries due to the lack of data (Östgaard 1965), their study covers the entire world by using the GDELT dataset. The main finding was a strong regionalism, and the important role of international news agencies in global news coverage.

There is a concern as to what extent the GDELT dataset is biased. Recently, a study that compares GDELT to Event Registry, another public news dataset, has been con-ducted (Kwak and An 2016). Two datasets showed a significant difference in terms of the number of news articles indexed and the active news media, and thus, authors suggest to use the dataset with the caution. However, at the same time, the study found that news geographies from two datasets were extremely similar, indicating that the overall trend of international news coverage seems to be consistent.

As a part of their vision to provide datasets of global society, GDELT recently released Visual Global Knowledge Graph (VGKG), indexing collected images and providing visual narratives of the world's news. Powered by the Google Cloud Vision API[1], each image turns into a set of textual features, such as topical focus of the image (what kinds of objects it includes), sentiment of people (whether people in the photo express joy, anger, sorrow, or surprise), identified locations, and even flagging of violent imagery, with confidence scores (0 to 1.0). VGKG datasets are publicly available from 1st January 2016.

## Datasets

We have downloaded VGKG from 00:00 1st January 2016 to 23:45 31st January 2016. We also have downloaded GKG during the same period. By using the URL, we match GKG and VGKG records so that we can augment VGKG data (properties of news photos) with GKG data (properties of news text), such as the tone of the news text.

Table 1 presents the summary of the data. In the table, top 500 indicates all the VGKG records from the top 500 news media websites listed by Alexa.com.

| Category | Number of records |
| --- | ---: |
| Top 500 (Eng) | 385,175 |
| Non-top 500 (Eng) | 2,070,655 |
| All (Eng) | 2,455,830 |

Table 1: Full Visual GKG dataset

The full data in Table 1 are the entire collection of the news articles written in English that GDLET monitors. It indexes not only popular news media but also news aggregators and even blogs. This mixture of news sources contains the practice of professional journalists and personal bloggers and makes it hard to focus on well-established practice.

To split them, we extract VGKG records of 7 news media considering the number of articles indexed in the VGKG collections (> 1,000) and media popularity (listed as Top 30 news websites according to Alexa.com): BBC, Bloomberg, CNN, Fox News, HuffingtonPost, Reuters, and Time. We exclude Yahoo News because our aim is to characterize each news outlet rather than a news portal that collects news from all the news outlets. Also, we exclude The Guardian because many of its indexed images are the placeholders of "Unauthorized" messages because it blocks direct access to the image. Table 2 summarizes 7 news media.

---

[1]https://cloud.google.com/vision/

| Origin | Media outlet | Record |
| --- | --- | ---: |
| US | The Huffington Post | 3,906 |
| UK | Reuters | 3,490 |
| UK | BBC | 3,249 |
| US | Fox News | 2,943 |
| US | CNN | 2,770 |
| US | Time | 1,545 |
| US | Bloomberg | 1,119 |

Table 2: Seven popular news media dataset

## Analysis of News Photos

### News Section and Photos

A typical newspaper is divided into multiple sections, such as 'politics', 'health', or 'entertainment'. Although this is coarse-grained, it indicates the topic of a news article. In the digital era, such section information is even captured from a URL to a news article.

Among the seven news media outlets, two news media outlets, CNN and Fox News, contain section names in the URL. For instance, we can recognize that the article is in the 'health' section from "http://edition.cnn.com/2015/12/31/health/auto-brewery-syndrome-dui-womans-body-brews-own-alcohol/". In this section, we examine how differently photos are used across the sections, and thus, we focus on CNN and Fox News.

Sections are divided into two categories: 1) subject-related and 2) regional sections. The first category is a subject of news. Of the above examples, 'politics' or 'technology', fall into this category. From CNN, we find nine subject-related sections: Politics, Opinion, Health, Entertainment, Investing, Media, Technology, Travel, and Living. From Fox News, we also find nine subject-related sections: Health, Politics, Entertainment, Opinion, Leisure, Tech, Science, Travel, and Life Style.

The other is regional categories–which country a news article is about or where the event happens. For example, in some news media, when something happens in Asia, that news appears in 'Asia' section no matter what happened. We find that CNN has a wider range of location sections, such as 'US', 'Asia', 'World', 'Middle East', 'Europe', and 'Africa', while Fox News has only 'US' and 'World'. We omit the sections that have less than 30 articles indexed.

We first investigate what frequently appears in news photos and how different they are across the sections. We use only labels with confidence scores higher than 0.8 computed by Google Cloud Vision API. Table 3 presents the top 10 most common objects in news photos in each of the subject-related sections.

Firstly, we observe that "Person" frequently appears in news photos except in the Travel section. For some sections, such as Politics, Opinions, and Entertainment, "Person" appears in more than 60% of all images. For comparison, we note that 40.5% of the photos from top 500 news media and 35.6% from non-top 500 news media contain "Person" or "People". In other words, we can see a high proportion of photos of people in news.

A set of labels such as Facial expression, Nose, Hair,

| Section | Top 10 common objects |
|---|---|
| Politics | Person (64.7%), People, Speech, Facial Expression, Athlete, Nose, Musician, Art, Performance, Profession |
| Opinions | Person (62.8%), People, Speech, Facial Expression, Man, Profession, Hairstyle, Musician, Athlete, Hair |
| Health | Person (27.6%), Macro Photograph, Handwriting, Text, Water Tower, Human action, Woman, Athlete, Infant, Shoe, Purple |
| Entertainment | Person (69.5%), Athlete, People, Musician, Hair, Hairstyle, Art, Guitarist, Facial expression, Man |
| Investing | Person (10.3%), Vehicle, Art, Transport, Clothing, Cartoon, Sports, Hairstyle, Hair, Airliner |
| Media | Person (50.0%), People, Athlete, Facial Expression, Art, Clothing, Speech, Hairstyle, Hair, Blond, Man |
| Technology | Person (30.1%), Art, Facial expression, Hairstyle, Hair, Cartoon, Eyebrow, Profession, Vehicle, Athlete |
| Travel | Art (8.5%), Vacation, Ship, Islet, Snow, Steppe, Vehicle, Beach, Passenger ship, Mode of transport |
| Living | Person (37.8%), Food, Sketch, Royal icing, Hair, Mascarpone, Facial expression, Woman, Child, Toppings |

Table 3: Top 10 objects appearing in news photos across the subject-related sections in CNN

Hairstyle, and Eyebrow are commonly found in Politics, Health, Entertainment, and Opinion. These labels indicate that among different ways to portray human, a close-up portrait is the most common across these sections. According to the study by Kress and Van Leeuwen (1996), a close-up portrait tends to be more intimate than a long shot to the viewers. Concerning that those sections are human-centered subjects, close-up photos may help them to portray people in the image to have a closer relationship with the readers.

In Health section, Macro photograph is the second most common object. Manual inspection finds they are magnified insects or medical pills. Also, Infant was depicted in the news articles about breastfeeding, cough syrup for children, etc. Infant was also found in the photos depicting general happiness (happy family with an infant).

In technology section, screenshots of software (e.g. Windows 10) are often labeled as cartoons.

Table 4 presents the top 10 most common objects in news photos across the region-related sections. Similarly to what we observe in subject-related sections, Person is the most common in region-related sections. It is interesting to note that US section includes Woman as one of the primary objects, but not other regions.

We observe that Asia and Europe are often associated with photos of Protest and Middle East with photos of Mil-itary and Soldier. This seems to be mainly due to the recent incidents regarding ISIS, Syria, and refugees in Europe and Asia. In the Africa section, Wilderness is one of the top 10 most common labels in the image. Golan reports that western mass media strengthen the portrayal of the third world by reporting war, poverty, famine, conflicts, violence and conflicts and lead to negative perception (Golan 2008). Our finding, objects related with food and wilderness in the Africa section, is aligned with such stereotypes of Africa and demonstrates the evidence of modern media using stereotypical representation of other part of worlds by large-scale image analysis.

| Section | Top 10 common objects |
|---|---|
| US | Person (29.8%), People, Hair, Athlete, Hairstyle, Man, Atmosphere, Vehicle, Woman, Night |
| Asia | Person (41.9%), People, Vehicle, Hairstyle, Hair, Man, Protest, Transport, Facial expression, Athlete |
| World | Person (39.2%), Vehicle, Athlete, Profession, Woman, Art, People, Facial expression, Boat, Logo |
| Middle East | Person (45.0%), People, Athlete, Protest, Transport, Clothing, Speech, Military, Soldier, Basketball player |
| Europe | Person (39.1%), People, Sea, Facial expression, Man, Marine biology, Nose, Protest, Eyebrow, Crowd |
| Africa | People (29.6%), Person, Vehicle, Food, Athlete, Drink, Coca cola, Soft drink, Painting, Wilderness |

Table 4: Top 10 objects appearing in news photos across the location-related sections in CNN

We now move our focus to Fox News and investigate what photos are used in their sections. Similarly to CNN, Person appears the most in photos in Fox News. However, the proportion of Person photos in Fox News is lower than in CNN– Person appears in 60.3% of CNN's news articles and 53.2% of Fox News's news articles. In politics, Speech and Facial expression frequently appear in politics. Also, the Leisure section has the lowest percentage of Person, as we observed for CNN.

In contrast, Tech news in Fox News is more focused on products (hence low percentage of Person). We find that the difference comes from the convention that CNN often reports more about people in the field, such as Mark Zuckerberg or Evan Wiliams, in news stories.

Unlike CNN, Fox News has Science as a separate section, not as a part of Technology and the most common object in Science is Person. When reporting scientific results, news media tends to have photos of scientists who conduct the corresponding research or people working together. Although it does not seem to be necessary, a news article, titled 'Bones of hunted mammoth show early human presence in Arctic[2],' had a photo of the scientist who excavated the mammoth carcass (Figure 1). It is the common practice in news industry to give credit to scientists or investors rather than scientific discovery itself.

We omit the region-related sections in Fox News because it has only two sections and does not show significant differences between them.

---

[2] http://goo.gl/1OpszE

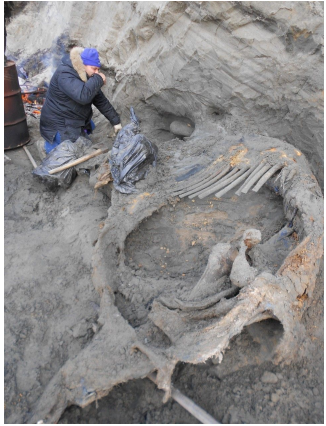

Figure 1: Labeled as 'Person' (0.853)

### Faces in News

In the previous section, we find that the vast proportion of news photos depict a person or people. We now provide in-depth analysis of how a person or people are portrayed, in terms of angles and emotionality, in news photos. In traditional settings, manual coding is unavoidable to conduct such study. However, thanks to Google Cloud Vision API which detects faces and the angles of the shot, and emotional attributes of the face (e.g., anger or joy), we can examine the representation of people in news photo at large-scale. We only consider the results of face detection with a high confidence score ($>0.8$).

First, we examine the seven popular news media individually to see how many faces are depicted in a news photo. Among photos containing people, 66.5% (Fox News) to 75.7% (BBC) of photos are single-person photos. Compared to the fractions of news articles mentioning one person (5.3% (Reuters) to 45.9% (BBC))[3], we see a strong tendency of using one-person photos in news articles even when the news articles mention more than one person. This tendency can be utilized to identify a key person in the corresponding news article that mentions multiple people.

We then examine the pattern of how faces are presented in news photos. VGKG measures the angle of the face in the photo by three metrics, widely used in photography: pan, roll, and tilt. Pan is a rotation on a horizontal plane of a camera. Roll is a rotation on a z-axis; when we take a photo in a landscape mode and portrait mode, our camera rolls. Lastly, tilt is literally a camera tilted up or down.

Figure 2 shows the distribution of how faces are presented in Fox News. The remaining five popular news media except Bloomberg have a similar distribution with Fox News, confirmed by Two-sample Kolmogorov-Smirnov test. Hence, we omit them due to lack of space. Bloomberg shows more flattened patterns of panning but we also omit it due to the lack of space.

---

[3]We use 'V2ENHANCEDPERSONS' field in GKG record to get information about people in the news article. The field lists all people referenced in the news article.

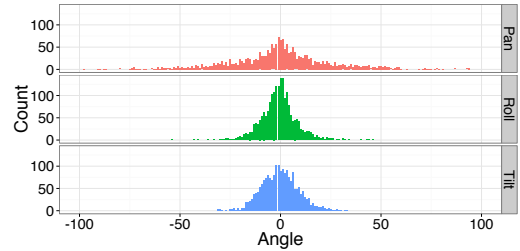

Figure 2: Face angles in Fox News

The common pattern is a bell-shape. The highest peak is observed around at zero, but not exactly zero. Journalists are most likely to use frontal view with a slight pan, roll, and tilt when selecting photos of people. Also, the range of panning (94 of Fox News to -98 of Fox News) is larger than that of rolling (53 of BBC to -63 of Huffington Post) and tilting (38 of Reuters to -38 of BBC). This is understandable because a highly rolling and tilting view of photos would simply look awkward.

The last feature that Google Cloud Vision API provides is detecting emotional attributes. Four emotional attributes, which are joy, anger, sorrow, and surprise, are measured in four levels: Very Unlikely (-2), Unlikely (-1), Likely (1), and Very Likely (2).

Among 11,127 faces collected from the seven popular news media, we find 2,740 (24.6%) faces having Likely or Very Likely values in one of the four emotional attributes. Among them, 2,665 (24.0%) faces express joy – 1 out of 4 faces in news photo is likely to express joy. The second highest proportion of emotional attributes is Sorrow, 0.6% (62 faces). Anger (13 faces) and Surprise (6 faces) are rare.

We also conduct the same experiment only with faces that very likely have emotional attributes. 21.1% faces have emotional attributes with Very Likely values – 1 out of 5 faces in a news photo is very likely to express joy. Almost all the faces are associated with Joy (2,336 faces). Sorrow decreases to 6 faces, Anger to 3 faces, and Surprise becomes zero faces. This shows that emotions other than joy are not likely to be expressed in news photos.

The results show a clear bias toward faces conveying positive emotion. However, this contrasts with the recent sentiment analysis of headlines–negative news headlines were the majority (more than 50% of news articles), followed by neutral and positive ones (Reis et al. 2015). This implies that, in the worst case, the expression of faces in the news photos may not be matched to the overall tone of the news articles. We examine the relationship between the tone of the news and the expression of the faces in the next section.

### Tone of Text and Smile in Photos

Does the emotion of the photo match with tone of the text? To answer to this question, we need to know 1) the fine-grained emotionality of the faces in the photo and 2) the tone of the text. For the first, we use another Face API developed by Microsoft Oxford Project[4] and add meta-information of images. Face API recognizes human faces and estimates the gender, age, and smile intensity (0 to 100) of the faces given a photo. We use smile intensity as a measure of the level of emotion of the photo. For the latter, we use GKG data which provides the average tone of the text. It ranges from -100 (extremely negative) to +100. We round the tone and make 201 bins for the tone. We count the number of articles that fall into each bin. For robust analysis, we filter out all the bins that have fewer than 100 articles. Then, we draw the box-plot of smile intensity for each bin.

---

[4]https://www.projectoxford.ai/face

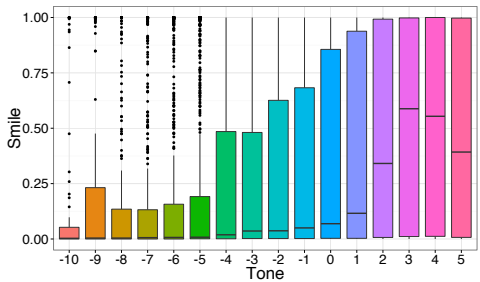

Figure 3: Relationship between the tone of the text and smile intensity of the photo

We can see an overall positive trend between the tone of the text and the smile intensity of the photo. We confirm that the positive tendency is statistically significant by computing Spearman correlation coefficient ($\rho$=0.225, p < 0.0001).

The alignment between smile intensity and the tone of the text is not only about whether it looks natural or awkward but also whether it actually helps understanding of readers. Stenberg et al. (Stenberg, Wiking, and Dahl 1998) found that happy faces accelerate the cognitive processing of positive words and slow down that of negative words through control experiments.

### Gender in News Photos

Researchers have long criticized the media for its stereotypical representation of gender. Previous studies have examined photos in newspapers and have drawn the same conclusions: 1) men outnumber women and 2) men and women are portrayed associated with particular roles (Brabant and Mooney 1986; Bridge 1997; Adams and Tuggle 2004). More recently, Rodgers et al. (2007) have found emotional stereotypes by gender in news photos–significantly more women than men were depicted as happy, calm, and submissive. In contrast, significantly more men than women were portrayed as sad, excited, and dominant.

To investigate the portrayals of men and women in news photos, we need to infer the gender of a person in the photo. As we describe in the previous section, we use Face API by MS Oxford Project to estimate the gender, age, and smile intensity of the faces in news photos.

Having gender of people in the news photos at hand, we now examine how many men or women appear in news photos. Figure 4 shows stark differences in the percentage of female photos across different sections within CNN. We find a significantly large proportion of female photos in the Living section, consistent with the results in previous studies that women appear more in homemaking or entertainment (Luebke 1989; Miller 1975; Rodgers and Thorson 2000). In the Health section, we also see more women than men in news photos. We also find quite a high proportion of women in the technology section, which is somewhat unexpected. By manual inspection, we find those news articles highlight a few specific female figures in technology domain, such as Sheryl Sandberg or Elizabeth Holmes. When campaigning the need of general coding education, high schools girls' photos were used.

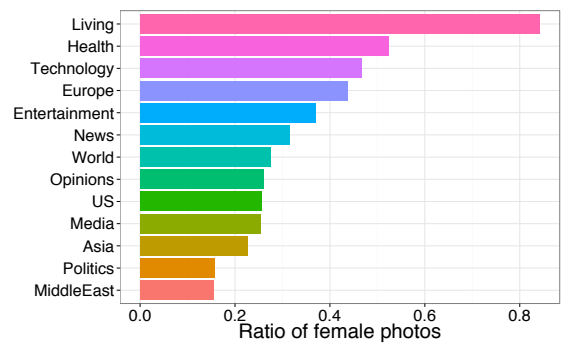

Figure 4: Ratio of female photos in each section (in CNN)

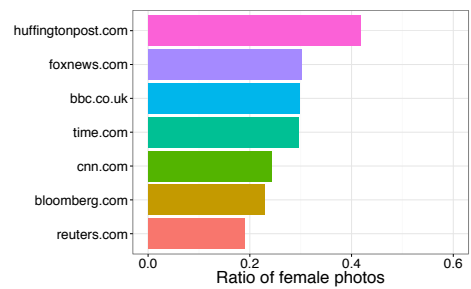

Figure 5: Ratio of female photos in each media outlet

In Figure 5, we show the ratio of female photos across the media outlets. Similar to previous studies, men outnumbers women (Luebke 1989) across all media outlets. We see the highest proportion of female photos from Huffington Post and the lowest one from Reuters. This will be related to the topics that each outlet covers more.

Figure 6 presents the intensity of smiles by gender in each news outlet. Women smile more than men in all seven popular news media–the smile intensity of females is stronger than that of males. Even for Reuters, which has the smallest gap of median smile intensity between male and female, the difference between male and female is statistically significant but the effect size is small (p=.0189, Z=2.3464, r=0.06).

LaFrance et al. find that, in general, women smile more than men through meta analysis of 162 research reports (LaFrance, Hecht, and Paluck 2003), while there are some differences across social or cultural contexts (Brody and Hall 2008). Also Rodgers et al. (2007) report that women are stereotyped as happier than men in news photos.

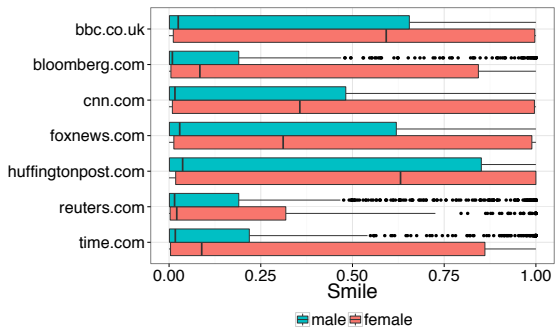

Figure 6: Relationship between gender and smile intensity

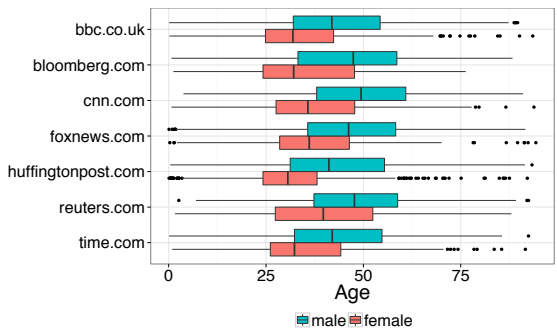

Figure 7: Relationship between gender and age

We then present the relationship between gender and age of human faces in news photos in Figure 7. For each of the news media, the difference in median age between the males and females is statistically significant by a Mann-Whitney's U test ($p < 0.0001$). Females in news photos generally look younger than males.

In summary, Figures 6 and 7 show how differently men and women are portrayed in news photos. Women smiles more and look younger than men in the news.

## News Photos in Politics

In this section we conduct an extensive case study of news photos in politics, focusing on portrayals of different presidential candidates in different news outlets.

News media have long been accused of bias during political campaigns. More recently, scholars have started to study bias in the visual images presented in news coverage. In particular, those studies have focused on how candidates are portrayed visually in presidential coverage (Waldman and Devitt 1998; Moriarty and Popovich 1991; Barrett and Barrington 2005). The results are mixed, and studies are limited by examining a handful of images without a systematic way and in-depth analysis (Goodnow 2010).

Goodnow (2010) then has systematically examined photos whether such bias is evident by adopting Kress and van Leeuwen (1996)'s method of social semiotics. The analysis of *Time* magazine's photo essays on the Democratic candidate during 2008 presidential election, showed that the levels of action, intimacy, organization, and support for two candidates are portrayed in disparate ways.

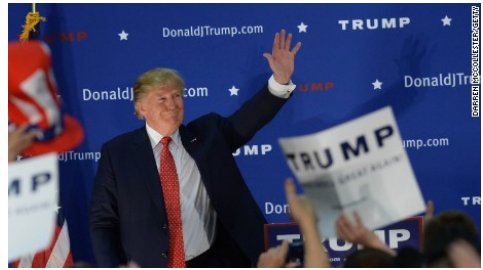

Figure 8: Labeled as 'Athlete' with high confidence score (0.941)

By manual inspection, we find that an image of a politician's victory pose is often labeled as athlete with a high confidence score, as in Figure 8[5]. Also, an image with a strong focus on the figure and a blurred background is often labeled as athlete because it is very similar to photos of in-game sports play. Despite this mislabeling, however, at the same time, we can use it as a proxy of photos conveying subtle positive nuances. Are specific politicians more likely to be labeled as an athlete?

We focus on five US politicians: two leading candidates for the Democratic nomination for 2016 presidential election, Hiliary Clinton and Bernie Sanders, and three leading candidates for the Republican nomination, Donald Trump, Marco Rubio, and Ted Cruz. We assess the implicit bias to them in both news media. As VGKG does not offer a feature to describe 'who' is in the photo, we infer it based on URL of the image. It roughly describes what is in the photo and is enough to decide whether Clinton or Sanders is in the photo. For example, we can extract Sanders from 'http://i2.cdn.turner.com/cnnnext/dam/assets/160117095426-bernie-sanders-gun-liability-corporations-reversal-sotu-00000000-large-169.jpg'. Even though GKG has the 'V2ENHANCEDPERSONS' field for each record, it is not straightforward to use because it returns a list of all persons referenced in the article.

Among 556 images labeled as 'person' in CNN, we find a candidate name from 355 images and multiple candidates from 73 images. We note that 8 out of 73 are marked as an athlete. Similarly, We find 219 images labeled as 'person' in Fox News and a candidate name from 88 images and multiple candidates from 6 images. Only one of them is marked as athlete. Since VGKG does not show which candidate is marked as an athlete among multiple candidates, we exclude these cases from the below analysis.

Table 5 presents the proportion of photos marked as an athlete for each candidate in CNN and Fox News. In both media, we find a strong tendency that Bernie Sanders is rarely labeled as an athlete, while other candidates have a similar proportion of athlete labels in CNN and Fox

---

[5]http://goo.gl/wCI98I

| Media | Clinton | Sanders | Trump | Rubio | Cruz |
|---|---|---|---|---|---|
| CNN | 17.6% | 1.9% | 19.5% | 25% | 20.5% |
| Fox | 22.7% | 0.0% | 20.0% | 33.3% | 6.3% |

Table 5: Proportion of politicians labeled as athletes

News. Considering the tightening race between Clinton and Sanders, there is no clear external reason why Sanders receives fewer athlete labels. Then, Rubio has the highest percentage in both media, while the margin is not big. The athlete-like photos might be related to biological factors–the age of candidates. Sanders is the oldest (74) among them, and Rubio is the youngest (44). While we do not know the image-making strategies of both candidates, it might be true that Sanders tries to project a calm and quiet attitude to stress his experience.

Another reason might be rooted in media bias toward a certain candidate. To investigate this, we check news media that have made endorsements before 1st February 2016 via "Newspaper endorsements in the United States presidential primaries, 2016" in Wikipedia[6]. We find 10 news media supporting Hilary Clinton and 4 news media supporting Bernie Sanders. The main media except for the New York Times have not yet expressed an endorsement and most of listed news media do not have enough number of articles indexed. Thus, we group the news media based on their support for the candidate (Hilary Clinton or Bernie Sanders) for the comparison.

| Candidate | Pro-Clinton media | Pro-Sanders media |
|---|---|---|
| Hilary Clinon | 28.8% (4/14) | N/A |
| Bernie Sanders | 0.0% (0/22) | N/A |

Table 6: Proportion of politicians labeled as an athlete in the news media that made endorsement for Clinton or Sanders

We extract the VGKG entries of those news media that have made endorsements to see how they portray the two candidates. In the pro-Clinton media, we can see that more photos of Sanders (22) are used than those of Clinton (14), but none of Sanders' photos is labeled as an athlete (Table 6). This is in agreement with the result from CNN and Fox News. Unfortunately, the pro-Sanders media are not as well indexed as VGKG, except Quad-City Times. While Quad-City Times has 2,072 VGKG records, all the images are the same, its logo; it blocks direct access to the image link.

Nevertheless, we demonstrate that it is possible to go beyond simple text analysis and conduct large-scale research on images and their nuances. Not only examining how many times presidential candidates are mentioned in news media, but how presidential candidates are reflected in news media. Such studies usually have been done with tedious manual image coding, but thanks to the deep-learning-based vision API, we are now able to capture the subtle context of photos. We hope to see more research into the bias toward a specific candidate to strengthen a certain portrayal through news photos by using the data collected over a longer period.

**Politician and Emotion**

We then aim to connect politicians and emotional attributes of their face photos. One of the features used for determining the favored candidate of Time is "Smile"–Goodnow (2010) found is that Obama smiles more than Clinton in photos. It is known that a smile gives a positive, non-threatening impression to viewers (Goffman 1979). Along this line of research, we also focus on the facial expression of candidate in the photos.

| Media | Clinton | Sanders |
|---|---|---|
| The Huffington Post | 21.9% (7/32) | 18.1% (8/44) |
| Reuters | 27.2% (3/11) | 0.0% (0/0) |
| BBC | 0.0% (0/0) | 0.0% (0/2) |
| Fox News | 17.4% (8/46) | 14.3% (1/7) |
| CNN* | 26.5% (9/34) | 0.0% (0/29) |
| Time | 20.0% (2/10) | 17.6% (3/17) |
| Bloomberg | 16.7% (1/6) | 41.7% (5/12) |

Table 7: 'Joy' faces for each candidate

Table 7 shows the proportion of faces expressing joy. Compared to 24.0% of faces expressing joy in the whole set of news photos, fewer photos of Clinton and Sanders expressing joy appear in news outlets.

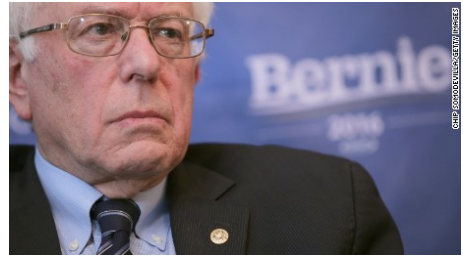

Figure 9: 'Sorrow' of Bernie Sanders in CNN

Another interesting finding is that in CNN we find two 'Sorrow' faces of Sanders as in Figure 9. No other candidates associate with sorrow, and Sanders does not associate with joy in CNN, even though his face appears 29 times.

| Media | Clinton | Sanders |
|---|---|---|
| Top 500 | 23.8% (226/951) | 14.6% (76/519) |
| Non-top 500 | 25.2% (953/3783) | 31.8% (841/2646) |
| Pro-Clinton | 21.4% (3/14) | 0.0% (0/4) |
| Pro-Sanders | N/A | N/A |

Table 8: 'Joy' faces for each candidate in the top 500 and non-top 500 news media

Along with candidates labeled as an athlete, we again find a somewhat different portrayal of Sanders in CNN, particularly compared to Clinton. For a fair comparison, we aggre-gate all the records from both top 500 and non-top 500 news media and examine how the two candidates are portrayed in the two datasets (Table 8). By using Fisher's exact test (the sample size of CNN is too small to use Chi-square test), we find that the proportion of 'joy' for Clinton is not different between CNN and non-top 500 news media ($p$=0.8439), but that of Sanders is different between CNN and non-top 500 news media ($p$=2.264e-05).

---
[6]https://goo.gl/nMCHAo

## Discussion

In this work, we have investigated more than two million news photos published during January 2016, indexed as Visual GKG powered by Google Cloud Vision API.

There are, of course, some limitations in this work. First, we only have looked into a one-month dataset. Data collected during a longer period will lead to more comprehensive and robust studies. Next, the high accuracy of Google Cloud Vision API is the key to the success of this kind of research. As Google Cloud Vision API shows the confidence score for every inference, we filtered out inferences with low scores. Also, one of our future research directions is to do comparative analysis between Google Cloud Vision API and Microsoft Project Oxford Vision API. Along with Face API, Microsoft Project Oxford also offers a wide range of vision API. The examination of how similar or different their descriptions of a given image are will help us to understand the accuracy of each API.

Next, some images were repeatedly used in multiple articles. We did not give special attention to them and treat them like other images, but it is possible for news media to strengthen some images of people by delivering the same image multiple times. Next, we recognize the photos of candidates by using URL, but face recognition technique can be used for this task. For instance, Microsoft Face API provides a feature called 'face verification' to check whether two faces belong to the same person or not. Finally, the size of the photo is not considered in this work at all. Like previous literature on the size of articles in the newspaper, the size of the photo should play some role to indicate the level of the importance of a given photo to readers.

Our aim was to show the great potential of deep learning-based vision technology for computational journalism. Like text mining has brought great advances in studies of textual contents of news, we believe that vision technology will open a new research area to understand news photos at large-scale. The new technology reduces time and effort for conducting such analysis with imagery content–it can easily tell whether the photo is a close-up shot or whether a person in the image is delivering a speech. We hope that our work will encourage researchers not only in computer science but also in other disciplines to have interests in large-scale analysis of news photos.